\documentclass[aps,prb,a4paper,superscriptaddress,twocolumn,showpacs,amsmath,amssymb,longbibliography]{revtex4-1}

\usepackage{graphicx,epsfig}
\usepackage[usenames]{color}

\usepackage[english]{babel}

\usepackage{csquotes}
\MakeOuterQuote{"}

\allowdisplaybreaks
\begin{document}

\newcommand{\cau}{\underline{c}^{\phantom{\dagger}}}
\newcommand{\ccu}{\underline{c}^\dagger}

\newcommand{\hu}{\underline{\hat{H}}}

\newcommand{\Dp}{\hat{\Delta}_{\text{p}}}
\newcommand{\Dh}{\hat{\Delta}_{\text{h}}}
\newcommand{\Deltap}{[\hat{\Delta}_{\text{p}}]}
\newcommand{\Deltah}{[\hat{\Delta}_{\text{h}}]}
\newcommand{\R}{\mathcal{R}}

\newcommand{\Rb}{\bar{\mathcal{R}}}
\newcommand{\Db}{\bar{\mathcal{D}}}

\newcommand{\Rh}{\hat{\mathcal{R}}}

\newcommand{\uA}{|\underline{A},Ri\rangle}
\newcommand{\uB}{|\underline{B},R'j\rangle}

\newcommand{\E}{\mathcal{E}}
\newcommand{\G}{\mathcal{G}}
\newcommand{\Lag}{\mathcal{L}}
\newcommand{\M}{\mathcal{M}}
\newcommand{\N}{\mathcal{N}}
\newcommand{\U}{\mathcal{U}}
\newcommand{\F}{\mathcal{F}}
\newcommand{\V}{\mathcal{V}}
\newcommand{\C}{\mathcal{C}}
\newcommand{\I}{\mathcal{I}}
\newcommand{\s}{\sigma}
\newcommand{\up}{\uparrow}
\newcommand{\dw}{\downarrow}
\newcommand{\h}{\hat{H}}
\newcommand{\himp}{\hat{H}_{\text{imp}}}
\newcommand{\g}{\mathcal{G}^{-1}_0}
\newcommand{\D}{\mathcal{D}}
\newcommand{\A}{\mathcal{A}}
\newcommand{\projs}{\hat{\mathcal{S}}_d}
\newcommand{\proj}{\hat{\mathcal{P}}_d}
\newcommand{\K}{\textbf{k}}
\newcommand{\Q}{\textbf{q}}
\newcommand{\T}{\tau_{\ast}}
\newcommand{\io}{i\omega_n}
\newcommand{\eps}{\varepsilon}
\newcommand{\+}{\dag}
\newcommand{\su}{\uparrow}
\newcommand{\giu}{\downarrow}
\newcommand{\0}[1]{\textbf{#1}}
\newcommand{\ca}{c^{\phantom{\dagger}}}
\newcommand{\cc}{c^\dagger}
\newcommand{\ga}{g^{\phantom{\dagger}}}
\newcommand{\gc}{g^\dagger}
\newcommand{\aaa}{a^{\phantom{\dagger}}}
\newcommand{\aac}{a^\dagger}
\newcommand{\bba}{b^{\phantom{\dagger}}}
\newcommand{\bbc}{b^\dagger}
\newcommand{\da}{d^{\phantom{\dagger}}}
\newcommand{\dc}{d^\dagger}
\newcommand{\fa}{f^{\phantom{\dagger}}}
\newcommand{\fc}{f^\dagger}
\newcommand{\ha}{h^{\phantom{\dagger}}}
\newcommand{\hc}{h^\dagger}
\newcommand{\be}{\begin{equation}}
\newcommand{\ee}{\end{equation}}
\newcommand{\bea}{\begin{eqnarray}}
\newcommand{\eea}{\end{eqnarray}}
\newcommand{\ba}{\begin{eqnarray*}}
\newcommand{\ea}{\end{eqnarray*}}
\newcommand{\dagga}{{\phantom{\dagger}}}
\newcommand{\bR}{\mathbf{R}}
\newcommand{\bQ}{\mathbf{Q}}
\newcommand{\bq}{\mathbf{q}}
\newcommand{\bqp}{\mathbf{q'}}
\newcommand{\bk}{\mathbf{k}}
\newcommand{\bh}{\mathbf{h}}
\newcommand{\bkp}{\mathbf{k'}}
\newcommand{\bp}{\mathbf{p}}
\newcommand{\bL}{\mathbf{L}}
\newcommand{\bRp}{\mathbf{R'}}
\newcommand{\bx}{\mathbf{x}}
\newcommand{\by}{\mathbf{y}}
\newcommand{\bz}{\mathbf{z}}
\newcommand{\br}{\mathbf{r}}
\newcommand{\Ima}{{\Im m}}
\newcommand{\Rea}{{\Re e}}
\newcommand{\Pj}[2]{|#1\rangle\langle #2|}
\newcommand{\ket}[1]{\vert#1\rangle}
\newcommand{\bra}[1]{\langle#1\vert}
\newcommand{\setof}[1]{\left\{#1\right\}}
\newcommand{\fract}[2]{\frac{\displaystyle #1}{\displaystyle #2}}
\newcommand{\Av}[2]{\langle #1|\,#2\,|#1\rangle}
\newcommand{\av}[1]{\langle #1 \rangle}
\newcommand{\Mel}[3]{\langle #1|#2\,|#3\rangle}
\newcommand{\Avs}[1]{\langle \,#1\,\rangle_0}
\newcommand{\eqn}[1]{(\ref{#1})}
\newcommand{\Tr}{\mathrm{Tr}}

\newcommand{\Vb}{\bar{\mathcal{V}}}
\newcommand{\Vd}{\Delta\mathcal{V}}
\def\P{P_{02}}
\newcommand{\Pb}{\bar{P}_{02}}
\newcommand{\Pd}{\Delta P_{02}}
\def\t{\theta_{02}}
\newcommand{\tb}{\bar{\theta}_{02}}
\newcommand{\td}{\Delta \theta_{02}}
\newcommand{\Rd}{\Delta R}

\title{Local bottom-up effective theory of non-local electronic interactions}

\author{Nicola Lanat\`a}
\affiliation{Department of Physics and Astronomy, Aarhus University, 8000,
Aarhus C, Denmark}

\date{\today} 

\begin{abstract}
A cardinal obstacle to understanding and predicting quantitatively the properties of solids and large molecules is that, for these systems, it is very challenging to describe beyond the mean-field level the quantum-mechanical interactions between electrons belonging to different atoms.
Here we show that there exists an exact dual equivalence relationship between the seemingly-distinct physical problems of describing local and non-local interactions in many-electron systems. This is accomplished using a theoretical construction analogue to the quantum link approach in lattice gauge theories, featuring the non-local electron-electron interactions as if they were mediated by auxiliary high-energy fermionic particles interacting in a purely-local fashion.
Besides providing an alternative theoretical direction of interpretation,
this result may allow us to study both local and non-local interactions on the same footing, utilizing the powerful state-of-the-art theoretical and computational frameworks already available.
\end{abstract}

\maketitle

\section{Introduction}
%
%
The phenomenon of strong electron correlations~\cite{Kent-Gabi} ---deeply related to what in chemistry is known as the ``multi-configurational problem''--- is widespread in materials with transition metals from the 3d series, lanthanides, actinides, as well as in organic matter.~\cite{Mott-organic2,Mott-organic3}
%
%
%
The need of explaining the spectacular emerging behaviours of strongly-correlated systems~\cite{Kent-Gabi}
---such as the Mott metal-insulator transition,~\cite{arresting_motion} high-temperature superconductivity~\cite{RevModPhys.78.17,Paglione2010} and magnetism,---
has led to the development of powerful theoretical frameworks, which are typically referred to as ``quantum embedding'' (QE) theories.~\cite{Kent-Gabi,QE-chan} Well-known examples are: dynamical mean field theory (DMFT),~\cite{LDA+U+DMFT,Held-review-DMFT,DMFT,dmft_book} multi-orbital generalizations of the Gutzwiller approximation, 
(GA)~\cite{Gutzwiller3,Gebhard,fab,Fang,lanata,Gmethod,Fang,Ho,Our-PRX,Ghost-GA} density matrix embedding theory (DMET),~\cite{DMET} the rotationally invariant slave boson theory (RISB~\cite{Kotliar-Ruckenstein,Georges,rotationally-invariant_SB,UO2} and the respective combinations of these approaches with MF methods
Moreover, new promising QE methodologies based on quantum-chemistry approaches are recently emerging.~\cite{QE-R-1,QE-R-2}
The key idea underlying all MF+QE methods consists in separating the system into: (1) a series of local fragments (called ``impurities''), which require a higher-level treatment due to the presence of strong-correlation effects (e.g., the d open shells of transition metals) and (2) their surrounding environment, which is treated at the mean field level.
%
%
The fundamental reason underlying the predictive power of the MF+QE methodologies is that they describe the \emph{local} electronic interactions beyond the mean-field level. Therefore, these theoretical frameworks capture the characteristic atomic energy scales emerging in strongly correlated matter, which are at the basis of many of the properties of these systems.~\cite{Kent-Gabi}
However, at present, the problem of describing beyond the mean-field level also the \emph{non-local} electron-electron interactions of realistic large molecules and solids is still very difficult. Indeed, this is a key limitation to our ability of understanding and simulating quantitatively strongly correlated systems, as the non-local interactions decay slowly with the inter-atomic distance and, in fact, they are often so large that they influence dramatically the electronic structure and generate new emerging phenomena, such as charge ordering.~\cite{CO1,CO2,CO3,CO4,CO5,CO6}
Remarkably, the non-local effects are of the utmost importance also in organic systems. A well-known example are the so-called ``London dispersion interactions,'' which affect the electronic structure of essentially all large condensed-phase systems,~\cite{VW1} including, e.g., the non-covalent bonds that determine the double-helical structure of DNA.~\cite{VW2}

Therefore, treating beyond the mean-field level both local and non-local interactions is very important.
This has stimulated intensive research and led to the
development of extensions of  DMFT~\cite{extendedDMFT1,extendedDMFT2,extendedDMFT3,extendedDMFT4,extendedDMFT5,extendedDMFT6} 
and the GA.~\cite{extendedGA1,extendedGA2,extendedGA3,extendedGA4,extendedGA5,extendedGA6,extendedGA7}
Nevertheless, the systematic inclusion of the non-local Coulomb interaction remains a serious challenge.


Here we 
derive a mathematically-exact reformulation of the problem,
where the non-local electronic interactions are replaced by local interactions with auxiliary fermionic degrees of freedom.
This result establishes a rigorous ``dual'' relationship between the seemingly-distinct physical problems of describing local and non-local interactions.
Furthermore, it may allow us to describe both of these effects
with the QE theoretical frameworks already available, in combination with the rapidly-evolving technological developments~\cite{QChem1,QChem2,QChem3,ML-DMFT,ML-mine,quantum-computation1,quantum-computation2,quantum-computation3,quantum-computation4,quantum-computation5,quantum-computation6,quantum-computation7,quantum-computation8}
for speeding up this type of calculations.

\section{Locality of interactions in effective theories}
When there are large energy scales that are well separated from the low-energy sector, the observables at one scale are not directly sensitive to the physics at significantly different scales. In some cases, such scale hierarchy constitute a great simplification, as the physics of the low-energy sector can be formulated in terms of an effective theory constructed in an exponentially-smaller Hilbert space. This perspective is typically referred to as ``top-down''. 
For example, 
the laws underlying the physics of ordinary solids, molecules and the whole chemistry are essentially encoded in quantum electrodynamics, whose formulation does not require to introduce particles such as the Higgs or W bosons. 
A complementary perspective is the so-called ``bottom-up'' approach, where one starts from a low-energy model and attempts to work up a chain of more and more ``fundamental'' effective theories consistent with the known low-energy physics, but valid also at higher energies.

A cardinal observation ---at the core of the present work--- is that low-energy effective theories 
can involve non-local interactions even if the original underlying theory is purely 
local.~\cite{non-loc-eff1,non-loc-eff2,non-loc-eff3,non-loc-eff4}
Here we are going to turn this problem 
into an advantage. In fact, we will show that it is possible to formulate
the physical Hamiltonian of a general 
multi-orbital extended Hubbard model ---containing
the ``troublesome'' non-local interactions present in all solids and molecules---  as the low-energy model of an underlying effective bottom-up fermionic theory with purely-local interactions.


\begin{figure}
\begin{center}
\includegraphics[width=8.4cm]{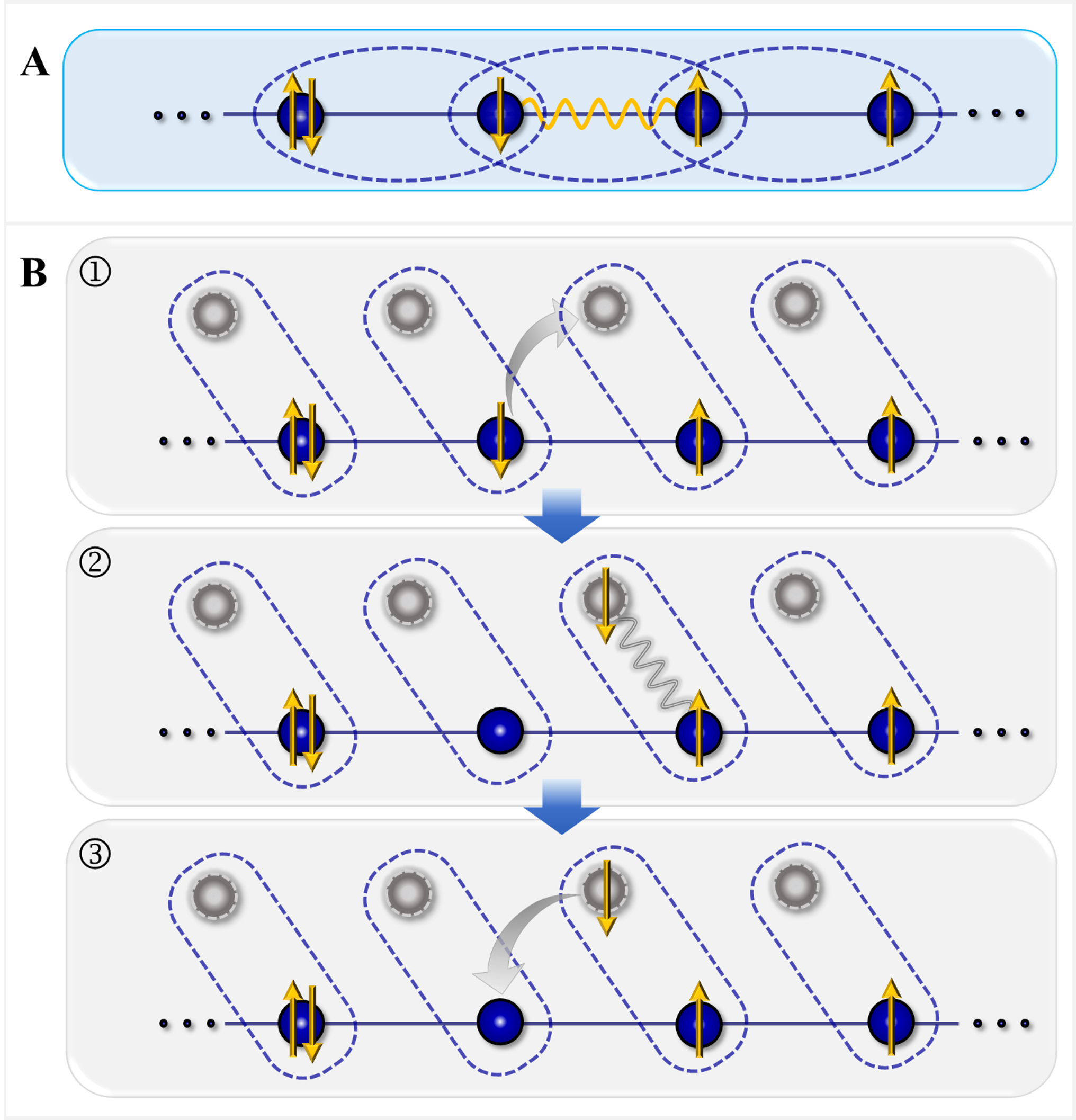}
\caption{Representation of the density-density interaction in the extended Hubbard model in dimension $D=1$. The blue bullets represent the physical electronic degrees of freedom and the gray bullets represent the ghost particles.
Panel A (Eq.~\eqref{hamgen}): The density-density interaction (yellow wavy line) is non-local.
Panel B (Eq.~\eqref{BU2}): The density-density interaction is effectively mediated  by the virtual processes 1,2,3, involving the ghost degrees of freedom. The interaction (gray wavy line) is local, i.e., it couples only ghost and physical degrees of freedom belonging to the same unit cell.
}
\label{Figure1}
\end{center}
\end{figure}

\section{Local bottom-up effective theory of the extended Hubbard model}
Before discussing realistic multi-orbital systems, let us
consider the periodic single-band extended Hubbard model on a $D$-dimensional hypercubic lattice:
\begin{align}
\h_{UV} = \h_{U}+\frac{V}{2}\sum_{\langle ij\rangle}
\hat{n}_{i}\hat{n}_{j}
\,,
\label{hamgen}
\end{align}
where 
$\h_U$ is the Hubbard Hamiltonian:
\begin{align}
\h_{U}=-t \sum_{\langle ij\rangle\sigma} \cc_{i\sigma}\ca_{j\sigma}
+U\sum_{i}\hat{n}_{i\uparrow}\hat{n}_{i\downarrow}
-\mu\sum_{i}\hat{n}_i\,;
\label{Hubbard}
\end{align}
the symbol $\langle ij\rangle$ indicates the summation over all nearest-neighbour pairs (so that each pair is counted twice);
$\cc_{i\sigma}$, $\ca_{i\sigma}$ are the annihiliation and creation operators of electrons of spin $\sigma\in\{\uparrow,\downarrow\}$ at the atomic site $i$; 
$\hat{n}_{i\sigma}=\cc_{i\sigma}\ca_{i\sigma}$ and $\hat{n}_{i}=\hat{n}_{i\uparrow}+\hat{n}_{i\downarrow}$.
The non-local operator proportional to $V>0$ is called ``density-density interaction''.~\cite{mainU} 

The key distinctive feature of the non-local interactions, such as the density-density terms in Eq.~\eqref{hamgen}, is that they make it impossible to partition the system into distinct subsystems coupled only by quadratic operators.
For example, the partitions enclosed by blue dashed lines in Fig.~\ref{Figure1}-A are not distinct, as they overlap with each other.
In many QE theories, this makes it very challenging to model accurately the coupling between the subsystems and their respective environments.
In particular, this is a well-known obstacle within many widely-used QE embedding methods (such as DMFT, RISB and the GA).
To solve this cardinal problem, here we design an effective theory satisfying the following conditions:
\begin{itemize}
\item[1.] Locality, i.e., the existence of a partition into finite subsystems coupled only quadratically.
\item[2.] Equivalence to Eq~\eqref{hamgen}, i.e., with the property of reproducing \emph{exactly} its physics  for all $U, t , \mu, V$.
\end{itemize}

\begin{figure}
\begin{center}
\includegraphics[width=8.4cm]{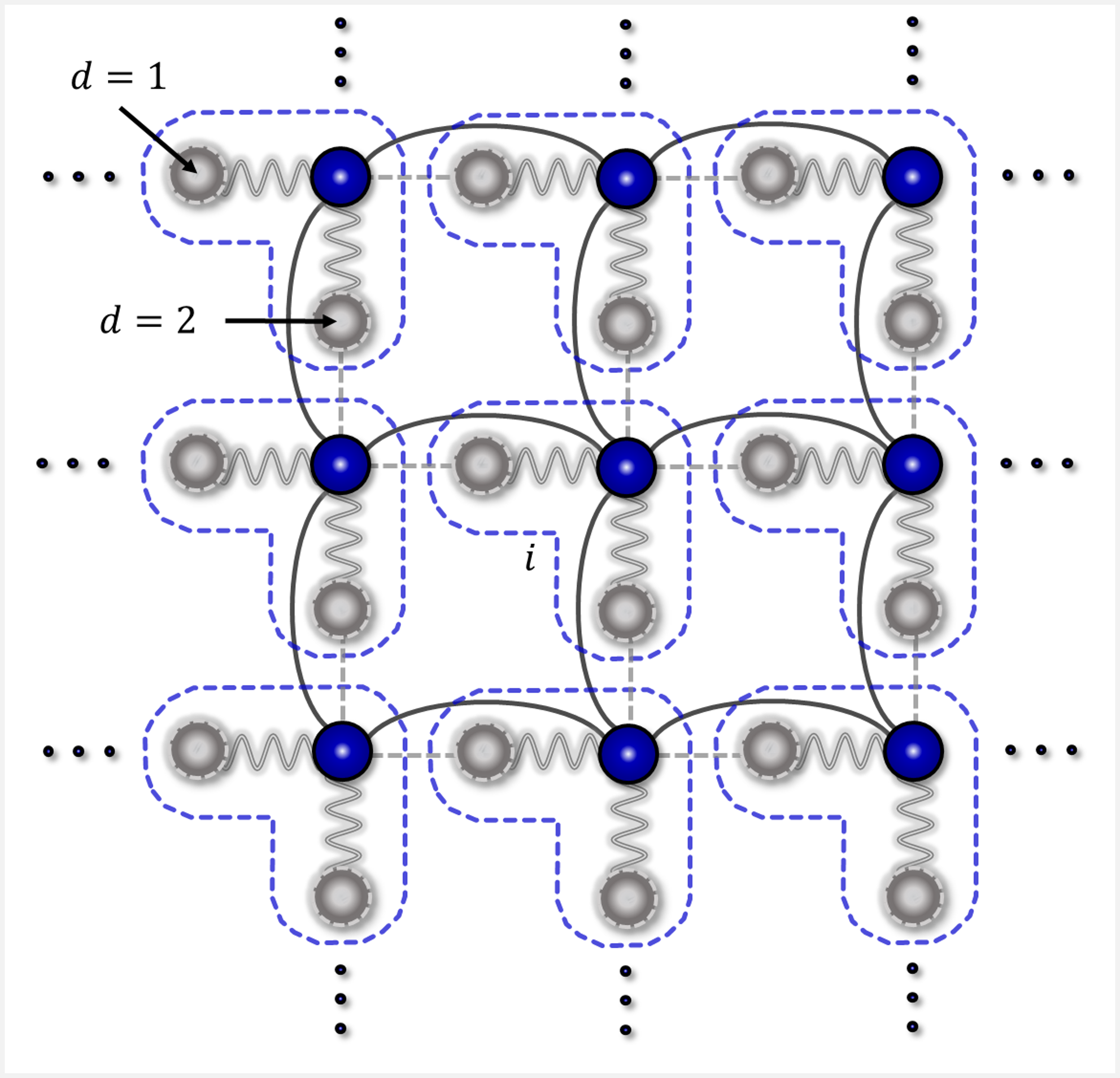}
\caption{
Representation of the effective bottom-up Hamiltonian
[Eq.~\eqref{BU2}] for a square lattice ($D=2$).
The blue bullets represent the physical electronic degrees of freedom and the gray bullets (placed on the corresponding lattice links) represent the ghost particles. 
The ghost modes placed on the horizontal links have label $d=1$, while the ghost modes placed on the vertical links have label $d=2$
The lattice unit cell, enclosed by dashed lines, have label $i$.
The density-density interactions between ghost and fermionic modes are indicated by gray wavy lines,
the hopping between physical modes is indicated by continue black lines,
the hopping between physical and ghost modes is indicated by gray dashed lines.
}
\label{Figure2}
\end{center}
\end{figure}

As we are going to show, such construction can be realized starting from the following Hamiltonian, represented in Fig.~\ref{Figure2} for dimension $D=2$:
\begin{align}
\h^{\gamma\tau}_{UV}&=\hat{H}_{U}+
\tau D\sum_{i}\hat{n}_{i}+
\gamma\tau
\sum_{i\sigma}\sum_{d=1}^{D}
\left(\cc_{i\sigma}\ga_{i+e_d d\sigma} \!+ \!\text{H.c}.\right)
\nonumber\\
&+\gamma^2
\sum_{i}\!\sum_{d=1}^{D}
\left(
\tau \, \hat{n}_{id}+V\, \hat{n}_{id}\hat{n}_{i}
\right)
\,,
\label{BU2}
\end{align}
where we have introduced $D$ auxiliary fermionic degrees of freedom $\gc_{id\sigma}$, $\ga_{id\sigma}$ for each unit cell $i$ and spin $\sigma$;
$\hat{n}_{id}=\sum_{\sigma}\gc_{id\sigma}\ga_{id\sigma}$ and $e_d$ is the $D$-dimensional vector with all entries equal to $0$ except for the $d$-th component, which is $1$.
From now on, we will refer to the auxiliary fermionic particles as the ``ghost'' degrees of freedom and to Eq.~\eqref{BU2} as the
``effective bottom-up theory''  of Eq.~\eqref{hamgen}.

The fact that Eq.~\eqref{BU2} is local (condition 1) stems from the fact that all interaction operators carry the same unit-cell label $i$. This is also shown in Fig.~\ref{Figure2}, where the subsystems indicated by the blue dashed lines interact only quadratically.
Note that the ghost modes 
(gray dots)
are associated with the links connecting the lattice sites (blue dots).
This structure is analogue to the quantum link approach to lattice gauge theories, where the fermionic fields are placed on the lattice sites, while the bosonic gauge fields are placed on the links.~\cite{Lattice-gauge-theories}
This analogy will be discussed further below.

Let us now focus on the equivalence to Eq.~\eqref{hamgen} (condition 2).
As we are going to show, for  $\tau\rightarrow\infty$ and $\gamma\rightarrow\infty$, $\h^{\gamma\tau}_{UV}$ reproduces exactly the physics of the extended Hubbard model [Eq.~\eqref{hamgen}].
In particular, this means that, for all physical observables $\hat{O}$ (constructed with $\ca_{i\sigma}$, $\cc_{i\sigma}$): 
\be
\lim_{\tau\rightarrow\infty}\lim_{\gamma\rightarrow\infty} 
\Av{\Psi_{\gamma\tau}}{\hat{O}}
= \Av{\Psi}{\hat{O}} 
\,,
\label{gs}
\ee
where $\ket{\Psi_{\gamma\tau}}$ is the ground state of Eq.~\eqref{BU2}
and $\ket{\Psi}$ is the ground state of Eq.~\eqref{hamgen}. 

Before demonstrating formally this fact, it is insightful to describe
intuitively the key physical concept underlying the construction
of Eq.~\eqref{BU2}:
The non-local interactions between the physical electronic modes $\ca_{i\sigma}$, $\cc_{i\sigma}$ (Fig.~\ref{Figure1}-A) can be viewed as if they were mediated by the ghost fermions (represented by the  operators $\ga_{id\sigma}$, $\gc_{id\sigma}$). Specifically, in $\h^{\gamma\tau}_{UV}$ the effect of $V$ is  the outcome of the following second-order sequence of processes (Fig.~\ref{Figure1}-B): (1) a  non-local hopping between physical and ghost modes, (2) a local density-density interaction between ghost and physical particles and (3) a second non-local hopping between physical and ghost modes.
%

Let us now prove this result mathematically. 
We define $P$ the projector over the ``physical'' space $\mathcal{V}_P$ generated only by $\ca_{i\sigma}$, $\cc_{i\sigma}$, i.e., where all ghost-particle occupation numbers are $0$. The projector over the auxiliary space generated by the eigenstates of $\hat{n}_g=\sum_{id}\hat{n}_{id}$ with eigenvalue $n_g\geq 1$ is $Q=1-P$.
From the Schr\"odinger equation for Eq.~\eqref{BU2}, we deduce  that the eigenstates of  Eq.~\eqref{BU2} satisfy the following equation:
\begin{equation}
\h_{\text{eff}}^{UV,\gamma\tau}[E_{\gamma\tau}]\,P\ket{\Psi_{\gamma\tau}}
=E_{\gamma\tau}\,P\ket{\Psi_{\gamma\tau}}\,,
\label{nonlinear}
\end{equation}
where $E_{\gamma\tau}$ is a generic eigenvalue of Eq.~\eqref{BU2} and:
\begin{widetext}
\begin{align}
\h_{\text{eff}}^{UV,\gamma\tau}[E_{\gamma\tau}]&=
P\h_{UV}^{\gamma\tau}P+P\h_{UV}^{\gamma\tau}Q\frac{1}{E_{\gamma\tau}-Q\h_{UV}^{\gamma\tau}Q}Q\h_{UV}^{\gamma\tau}P
\nonumber\\
&=\h_{U}+\tau D\sum_{i}\hat{n}_{i}+
\sum_{id\sigma}
\cc_{i\sigma}
\frac{\tau^2}{\gamma^{-2}E_{\gamma\tau}- 
\gamma^{-2}\big[\h_{U}+\tau D\, \hat{n}_{i+e_d}
\big] 
+
\left(
\tau +V \hat{n}_{i}
\right)
}
\ca_{i\sigma}
\,.
\label{intermediate_step}
\end{align}
\end{widetext}


Note that Eq.~\eqref{nonlinear}
---with $\h_{\text{eff}}^{UV,\gamma\tau}[E_{\gamma\tau}]$ given by the
first line of Eq.~\eqref{intermediate_step}--- is an exact identity valid for all Hamiltonians, see Ref.~\onlinecite{Effective_Hamiltonians}.
The second line of Eq.~\eqref{intermediate_step} is obtained by using that, for our specific Hamiltonian, each term of
$P\h_{UV}^{\gamma\tau}Q$ can rise the occupation number of only one of the ghost modes (from $0$ to $1$).

It is important to note that, at any finite 
$\gamma$, the effective Hamiltonian [Eq.~\eqref{intermediate_step}] depends itself on $E_{\gamma\tau}$.
To understand how Eq~\eqref{intermediate_step} simplifies for
$\gamma\rightarrow\infty$ (at fixed $\tau>0$), we need to estimate
how the eigenvalues $E_{\gamma\tau}$ of Eq.~\eqref{BU2} behave in this limit.
As we are going to show, the spectra of Eq.~\eqref{BU2} is divided in 2 sectors:
a the low-energy (physical) sector, such that:
\be
\lim_{\gamma\rightarrow\infty}\gamma^{-2}E_{\gamma\tau}=0
\,,
\label{Ebehavior}
\ee
and a high-energy (auxiliary) sector, with eigenvalues diverging as $\tau\gamma^2$. 
To prove this fact, starting from Eq.~\eqref{BU2}, we note that
$\gamma^{-2}\h^{\gamma\tau}_{UV}$ can be expressed as follows:
\be
\gamma^{-2}\h^{\gamma\tau}_{UV}=
\hat{h}^\tau_{UV}+
\hat{r}^\tau_{UV}(\gamma)\,,
\label{inverse-expansion}
\ee
where:
\begin{align}
\hat{h}^\tau_{UV}&=
\sum_{i}\!\sum_{d=1}^{D}
\left(
\tau \, \hat{n}_{id}+V\, \hat{n}_{id}\hat{n}_{i}
\right)
\,,
\\
\hat{r}^{\tau}_{UV}(\gamma)&=
\gamma^{-1}\tau
\sum_{i\sigma}\sum_{d=1}^{D}
\left(\cc_{i\sigma}\ga_{i+e_d d\sigma} + \text{H.c}.\right)
\nonumber\\&
+\gamma^{-2}\left(
\hat{H}_{U}+\tau D\sum_{i}\hat{n}_{i}
\right)\,.
\end{align}
%

We note that $\hat{h}^\tau_{UV}$ assigns an energy cost $\propto\tau>0$ to all unphysical configurations (with non-zero occupied ghost modes), while its ground space coincides with the physical subspace $\mathcal{V}_P$ (where, by definition, all ghost modes are empty).
Since $\hat{r}^{\tau}_{UV}(\gamma)$ 
vanishes for $\gamma\rightarrow\infty$,
the low-energy eigenvalues $\gamma^{-2}E_{\gamma\tau}$
of Eq.~\eqref{inverse-expansion}
satisfy Eq.~\eqref{Ebehavior}. Instead, $E_{\gamma\tau}$ diverges as $\tau\gamma^2$ for the unphysical states.
For the same reason, the low-energy eigenstates
$\ket{\Psi_{\gamma\tau}}$ become equal to $P\ket{\Psi_{\gamma\tau}}$ in this limit.

By substituting Eq.~\eqref{Ebehavior} in Eq.~\eqref{intermediate_step}, we deduce that, in the low-energy sector, Eq.~\eqref{nonlinear}
reduces to an ordinary (energy-independent) Schr\"odinger equation 
for $\gamma\rightarrow\infty$,
with respect to the following effective Hamiltonian:
\begin{align}
\h_{\text{eff}}^{UV,\tau}=
\h_{U}+\tau D\sum_{i}\hat{n}_{i}-
\sum_{id\sigma}
\cc_{i\sigma} 
\frac{\tau^2}{\tau +V \hat{n}_{i+e_d}}
\ca_{i\sigma}
\,.
\label{heff-gammalim}
\end{align}

Let us now evaluate the limit of Eq.~\eqref{heff-gammalim} for $\tau\rightarrow\infty$.
We note that, if $\tau$ is sufficiently large, the following equation holds:
\be
\frac{\tau^2}{\tau +V \hat{n}_{i+e_d}}
=\tau\left[
1-\frac{V}{\tau}\hat{n}_{i+e_d}
+ \sum_{l=2}^\infty \left(-\frac{V}{\tau}\hat{n}_{i+e_d}\right)^l
\right]\,.
\label{geometric}
\ee
In fact, since the maximum eigenvalue of $\hat{n}_{i+e_d}$ is $2$, the geometric series is guaranteed to converge for $\forall\,\tau>2V$.
By substituting Eq.~\eqref{geometric} in Eq.~\eqref{heff-gammalim}, 
we obtain:
\begin{align}
\h_{\text{eff}}^{UV,\tau}=\h_{UV}+\hat{o}_{V}(\tau)
\,,
\label{heff-taulim}
\end{align}
which coincides with Eq.~\eqref{hamgen} up to a perturbation of order
$\hat{o}_{V}(\tau)\sim V^2\tau^{-1}$, thatt  vanishes for $\tau\rightarrow\infty$.

In summary, for $\gamma\rightarrow\infty$ the states with non-zero ghost occupations are gapped out (as their energy diverges as $\tau\gamma^2$). On the other hand, they mediate the desired non-local density-density interaction of Eq.~\eqref{hamgen} within the physical space, by means of second-order (virtual) processes.
At finite $\tau$, these virtual processes generate also undesired additional interactions (because of the subleading terms of order $\geq 2$ in the geometric series [Eq.~\eqref{geometric}]).
But all spurious terms vanish as $2V/\tau^2$ for $\tau\rightarrow\infty$, while the desired density-density interaction remains.
In other words, the spurious terms associated with deviations from the extended Hubbard model are suppressed by a factor $\propto (V/\tau)^2$ with respect to the density-density interaction.
Therefore, in these limits, the low-energy sector of Eq.~\eqref{BU2} reproduces \emph{exactly} the physics of Eq.~\eqref{hamgen}, $\forall\,U,t,\mu,V$.

We point out that, in the last
step in Eq~\eqref{intermediate_step}, we used the fact  
that the ghost degrees of freedom in Eq.~\eqref{hamgen}
are placed on the links (analogously to
lattice gauge theories.~\cite{Lattice-gauge-theories})
In fact, linking a ghost fermion to multiple physical modes would generate additional second-order processes ---and, in turn, additional  
long-range interactions not present in the original extended
Hubbard Hamiltonian [Eq.~\eqref{hamgen}].
This explains why the non-local interactions along different directions in $\h_{UV}^{\gamma\tau}$ have to be mediated by distinct ghost particles, identified by the label $d$.
Note also that, since the ghost fermions are placed on the links, Eq.~\eqref{BU2} preserves the translational invariance of the system $\forall\,\gamma,\tau$ ---as opposed to classic cluster approaches,~\cite{CDMFT-Kotliar,CDMFT-Jarrell,CDMFT-Senechal,CDMFT-Potthoff,CDMFT-Lichtenstein} where the problem of breaking artificially the translational symmetry cannot be avoided.

\begin{figure}
\begin{center}
\includegraphics[width=8.4cm]{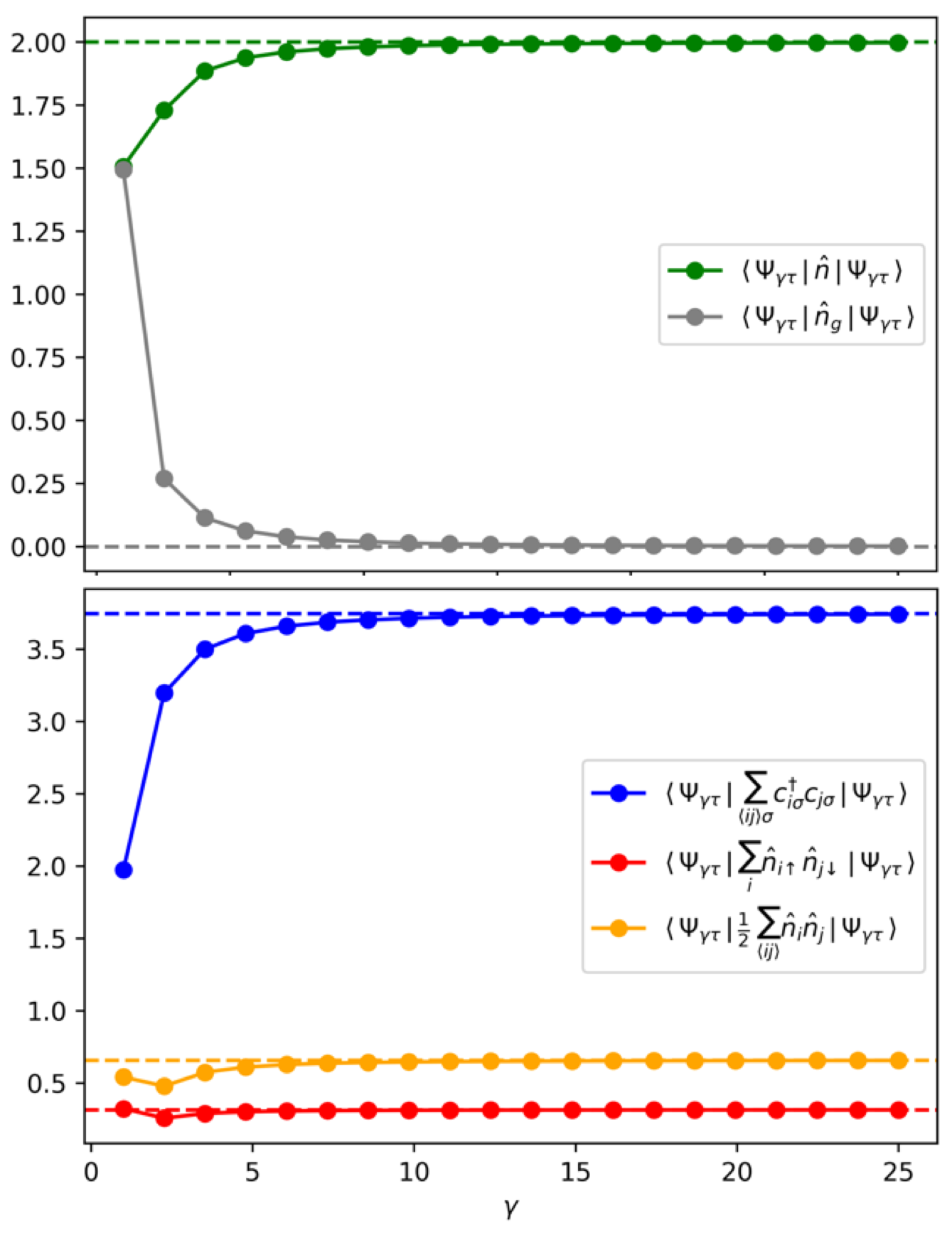}
\caption{Solution of the 4-sites single-band extended Hubbard model at half filling for $U=3$ and $V=1.5$, where $t$ is set as the energy unit.
The dotted lines indicate quantities calculated directly from Eq.~\eqref{hamgen}, while the bullet points are quantities calculated using the
effective bottom-up Hamiltonian [Eq.~\eqref{BU2}], for different values of $\gamma$
and ${2V}{\tau^{-1}}=10^{-2}$.
Top panel: Physical occupation (green) and ghost occupation (gray).
Bottom panel: Hopping operator (blue), double occupancy (red) and density-density operator (orange).
}
\label{Figure3}
\end{center}
\end{figure}

To illustrate the implications of our result in a minimal setting,
in Fig.~\ref{Figure3} we present numerical calculations of a half-filled 1-dimensional extended Hubbard model $\h_{UV}$ consisting of 4 physical sites.
Note that, for this relatively small system, both the original Hamiltonian and the corresponding effective Hamiltonian
$\h^{\gamma\tau}_{UV}$ can be diagonalized exactly.
%
As an example, we set $t$ as unit of energy, $U=3$ and $V=1.5$.
In all calculations we set the parameter of the geometric expansion
in Eq.~\eqref{geometric} to ${2V}{\tau^{-1}}=10^{-2}$.
In the top panel is shown the behavior, as a function of $\gamma$, of the physical occupancy $\Av{\Psi_{\gamma\tau}}{\hat{n}}$ (green line) and of the occupation of the ghost modes
$\Av{\Psi_{\gamma\tau}}{\hat{n}_d}$ (gray line).
In the bottom panel are shown the corresponding expectation values for a few
physical observables. 
Consistently with Eq.~\eqref{gs},
the numerical calculations confirm that the 
``spurious'' charge transfer between physical and ghost degrees of freedom vanishes 
for large $\gamma$, where the ghost modes are gapped-out and, therefore, the corresponding occupancy vanishes.
Furthermore, all expectation values converge to the correct limit for $\gamma\rightarrow\infty$, as expected.
The exact-diagonalization calculations were performed using the open-source software ``OpenFermion''.~\cite{OpenFermion}
\subsection*{Generalization to multi-orbital systems}
The procedure utilized above within the context of the single-band extended Hubbard model can be straightforwardly generalized to realistic multi-orbital Hamiltonians. For example, let us consider the following $D$-dimensional system:
\begin{align}
\h_{UV} = \h_{U}
+\frac{1}{2}\sum_{<ij>} 
\sum_{\alpha\beta\gamma\delta=1}^{\nu}
{V}_{\alpha\beta\gamma\delta}\,
\cc_{i\alpha}\ca_{i\beta}\cc_{j\gamma}\ca_{j\delta}
\,,
\label{hamgengen}
\end{align}
where
$\h_U$ is any multi-orbital Hamiltonian with purely-local interactions,
the labels $\alpha,\beta,\gamma,\delta$ represent both orbital and spin degrees of freedom and the coefficients $V_{\alpha\beta\gamma\delta}$ parametrize a generic density-density operator ---which is typically the largest contribution to non-local interactions in real solids and molecules.~\cite{mainU}
It can be readily shown that the effective bottom-up theory of Eq.~\eqref{hamgengen} is the following:
\begin{align}
\h_{UV}^{\gamma\tau}&=\h_{U}+
\tau D\sum_{i}\hat{n}_{i}
+\gamma\tau
\sum_{id\alpha} 
\left(\cc_{i\alpha}\ga_{i+e_{d}d\alpha}+ \text{H.c.}\right)
\nonumber\\
&+\gamma^2
\sum_{id} 
\left[
\tau \, \hat{n}_{id}+
\sum_{\alpha\beta\gamma\delta}
V_{\alpha\beta\gamma\delta}\,
\gc_{id\alpha}\ga_{id\beta}
\cc_{i\gamma}\ca_{i\delta}
\right]
\,,
\label{heffgen}
\end{align}
%
where:
\begin{align}
\hat{n}_{i}&=\sum_{\alpha}\cc_{i\alpha}\ca_{i\alpha}
\\
\hat{n}_{id}&=\sum_{\alpha}\gc_{id\alpha}\ga_{id\alpha}
\,.
\end{align}
Note that, as for the single-orbital case, $\h_{UV}^{\gamma\tau}$
becomes equivalent to $\h_{UV}$
for ${\gamma\rightarrow\infty}$ and ${\tau\rightarrow\infty}$
(where, as before, the limit for ${\gamma\rightarrow\infty}$ is taken first).

We point out that the theoretical construction derived in this work is not restricted to hypercubic periodic lattices, but can be extended to systems with arbitrary structures (also without translational symmetry).
Furthermore, also the Coulomb interactions beyond first nearest neighbours can be described, in a similar fashion, by introducing longer-range hopping operators between physical and ghost degrees of freedom.
Our approach may be applicable also for studying real crystals and molecules, e.g., using the interfaces already available for performing this type of calculations.~\cite{Haule10,HF+QE-1,Fang}
In fact, treating systematically beyond the mean-field level also the dominant (e.g., first-nearest-neighbor) contributions of the non-local interactions may improve substantially the accuracy of these methods.
In this respect, applications in combination with HF+QE frameworks~\cite{HF+QE-1} (which do not require to introduce any adjustable parameters) are particularly appealing.

\section{Conclusions}
In summary, we have shown that all multi-orbital systems 
represented by Eq.~\eqref{hamgengen}
---which includes all local and the largest non-local contributions of most real solids and molecules~\cite{mainU}---
can be equivalently described by solving higher-dimensional fermionic systems
with purely local interactions, see Eq.~\eqref{heffgen}.
This exact result is
analogue to the quantum link approach to lattice gauge theories, where the fermionic fields are placed on the lattice sites, while the bosonic gauge fields are placed on the links.~\cite{Lattice-gauge-theories}
The dual relationship between non-local and local interactions, demonstrated here,
provides us with a rigorous alternative direction of interpretation.
Furthermore, the possibility of using auxiliary fermions for decoupling the non-local interactions may allow us to develop new methods for including these effects in practical calculations ---complementary to the available computational frameworks. In fact, many of the existing state-of-the-art QE approaches are based on the so-called Hubbard-Stratonovich transformation,~\cite{extendedDMFT3} which utilizes auxiliary bosons, instead of fermions.
Computationally, the main cost of performing calculations of our dual model would be that introducing additional fermionic degrees of freedom increases the dimension of the EH.
However, within QE methods such as DMET and the GA (where the bath of the EH is as large as the impurity~\cite{Our-PRX,DMET}), applications to models with up to 3 correlated orbitals per atom may be already feasible using exact diagonalization or state-of-the-art quantum chemistry methods~\cite{QChem1,QChem2,QChem3} as impurity solvers.
Furthermore, thanks to the rapidly-evolving technological developments based on  machine learning~\cite{ML-DMFT,ML-mine} and quantum-computing,~\cite{quantum-computation1,quantum-computation2,quantum-computation3,quantum-computation4,quantum-computation5,quantum-computation6,quantum-computation7,quantum-computation8}
it may soon become possible to
apply this framework also to arbitrary d-electron and f-electron systems.

\section{Acknowledgements}
We thank Gabriel Kotliar, Garry Goldstein, Ove Christiansen, Vladimir Dobrosavljevi\'c, and Yongxin Yao for useful discussions.
We gratefully acknowledge funding from VILLUM FONDEN through the Villum Experiment project 00028019 and the Centre of Excellence for Dirac Materials (Grant. No. 11744). We also thank support from the Novo Nordisk Foundation through the Exploratory Interdisciplinary Synergy Programme project NNF19OC0057790.


%

\end{document}